\documentclass{article}
\usepackage{float}
\usepackage{placeins}
\usepackage{color}
\usepackage{soul}

\usepackage{spconf,amsmath,graphicx}
\usepackage{algorithm}
\usepackage{amssymb}
\usepackage{booktabs}
\usepackage{algpseudocode}
\usepackage{hyperref}

\title{Neural Network Augmented Kalman Filter for Robust Acoustic Howling Suppression}
\name{ \begin{tabular}{c} Yixuan Zhang$^{1*}$, Hao Zhang$^{2}$, Meng Yu$^{2}$, Dong Yu$^{2}$ \end{tabular}\thanks{$^{*}$This work was done during an internship at Tencent AI Lab.}}
\address{$^{1}$The Ohio State University, Columbus, OH, USA\\ $^{2}$Tencent AI Lab, Bellevue, WA, USA \\ }
\begin{document}
\ninept
\maketitle
\begin{abstract}
Acoustic howling suppression (AHS) is a critical challenge in audio communication systems. In this paper, we propose a novel approach that leverages the power of neural networks (NN) to enhance the performance of traditional Kalman filter algorithms for AHS. Specifically, our method involves the integration of NN modules into the Kalman filter, enabling refining reference signal, a key factor in effective adaptive filtering, and estimating covariance metrics for the filter which are crucial for adaptability in dynamic conditions, thereby obtaining improved AHS performance. As a result, the proposed method achieves improved AHS performance compared to both standalone NN and Kalman filter methods. Experimental evaluations validate the effectiveness of our approach.

% The NN aids in refining reference signal estimation, a key factor in effective adaptive filtering, and covariance estimation which is crucial for adaptability in dynamic conditions.

\end{abstract}
\section{Introduction}
\label{sec:intro}

% (What is acoustic howling, why is it importance to suppress it.)
Acoustic howling is a phenomenon that frequently arises in audio systems such as karaoke systems, and public address systems where the amplified sound from the loudspeaker is captured by the microphone and subsequently re-amplified recursively. This creates an internal positive feedback loop within audio systems, leading to an unpleasant howling sound that reinforces specific frequency components \cite{waterhouse1965theory, van2010fifty, loetwassana2007adaptive} which not only jeopardizes the proper functioning of equipment but also poses potential risks to human auditory health.
% (Briefly introduce existing methods for AHS: gain control, notch filter, AFC. Mention that AFC is better than other methods since .... The limitation of Kalman filter based AFC: sensitive to control parameters, not able to handle nonlinear distortions)
Various techniques have been explored to suppress howling including gain control \cite{schroeder1964improvement}, notch filter \cite{gil2009regularized, waterschoot2010comparative}, and adaptive feedback control (AFC) \cite{waterschoot2010comparative, joson1993adaptive, albu2018hybrid}. Notably, AFC methods leverage adaptive filters like the Kalman filter to suppress howling through continuous adjustments of filter coefficients driven by iterative feedback. 
The real-time adaptation of AFC methods breaks the positive feedback loop and results in better AHS performance. However, such methods have shown sensitivity to control parameters and cannot effectively manage the nonlinearity introduced by amplifiers and loudspeakers.

% (NN has been recently proposed for addressing AHS: introduce deepMFC, deepAHS and HybridAHS. Mention the shortcomings of NN based methods: mismatch problem, distortion problem in the enhanced signals.)

In recent years, inspired by acoustic echo cancellation (AEC) studies \cite{zhang2018deep, zhang2022neural}, deep-learning-based approaches have been explored to solve the AHS problem. Gan et al. \cite{gan2022howling} trained a deep neural network (DNN) in the time-frequency domain to suppress howling noise from speech signals. In \cite{chen2022neural}, a model based on a convolutional recurrent neural network is introduced for howling detection for real-time communication scenarios. In \cite{zheng2022deep}, a deep learning framework, called DeepMFC, is introduced to address marginal stability issues of acoustic feedback systems.  
While existing approaches seem promising, the data used for training is generated offline in a closed-loop system without AHS processing. This leads to a mismatch during streaming inference, as AHS processing is continuously integrated, influencing its input stream. To improve the mismatch issue and ease training, DeepAHS \cite{zhang2023deep} leverages the teacher-forcing strategy and demonstrates superior performance compared to other approaches. HybridAHS \cite{zhang2023hybrid} which incorporates the Kalman filter by augmenting its output as network input further improves AHS performance. Nevertheless, although both methods address the training concern and exhibit superiority, the discrepancy between training and real-time streaming inference still exists.
% Moreover, dealing with the distortion introduced into the enhanced signals presents a notable challenge in these approaches.
% (Introduce the motivation of this study: Leverage the advantages of both NN and Kalman. Briefly introduce previous studies such as NLMSNet, KalmanNet. What we proposed in this paper.)

NN-augmented adaptive filtering approaches which potentially introduce less distortion have been explored in the context of AEC. Deep Adaptive AEC \cite{zhang2022deep} employs NN modules to estimate the nonlinear reference and the step size in the normalized least mean square (NLMS) algorithm, which shows improved performance compared to fully DNN-based baselines in time-varying acoustic environments. Our prior study \cite{zhang2023kalmannet} integrates NN modules into a frequency-domain Kalman filter (FDKF) for estimating the nonlinear reference signal and a nonlinear transition function, which improves the performance of FDKF significantly and outperforms the NLMS-based Deep Adaptive AEC model. It is worth noting that our prior experiments show that exclusively employing NNs to estimate Kalman filter components does not necessarily yield performance improvements. However, leveraging NNs to estimate absent or approximated components within the Kalman filter algorithm has demonstrated considerable improvements, which further motivates our continued explorations in the field of AHS.

% which are crucial for effective adaptive filtering and the ability to handle dynamic changes in the acoustic environment.
% (Details of the proposed method; Briefly summarize the advantages of the proposed method.) 

In this study, we introduce NeuralKalmanAHS, an NN augmented Kalman filter for AHS. The proposed model incorporates NN modules into the frequency-domain Kalman filter, optimizing reference signal refinement and covariance matrix estimation. NeuralKalmanAHS is trained in a streaming mode that aligns with the streaming inference framework detailed in \cite{zhang2023deep} which evaluates AHS models in recurrent and real-world settings, thus eliminating potential mismatch issues. Furthermore, we employ a howling detection strategy during training to ensure model convergence, allowing successful model training even in challenging acoustic howling scenarios. 
Ablation studies indicate that streaming training ensures robustness of NeuralKalmanAHS against acoustic howling, even with lightweight models focused solely on covariance estimation, while reference signal refinement substantially boosts performance.
Experimental results show that the proposed NeuralKalmanAHS effectively suppresses howling noise with less distortion, demonstrating remarkable stability in challenging scenarios, and outperforming strong baseline methods.
\footnote{\label{note:demo}Demos are available in \url{https://yixuanz.github.io/NeuralKalmanAHS}.}

The remainder of this paper is organized as follows. Section \ref{sec:ahs} introduces the problem formulation of acoustic howling and the frequency-domain Kalman filter. Section \ref{sec:method} presents the proposed NeuralKalmanAHS. Section \ref{sec:exp} and Section \ref{sec:results} describe the experimental setup and evaluation results, respectively. Section \ref{sec:conclusion} concludes the paper.

\begin{figure*}[]
\centering
     \includegraphics[width=1.55\columnwidth]{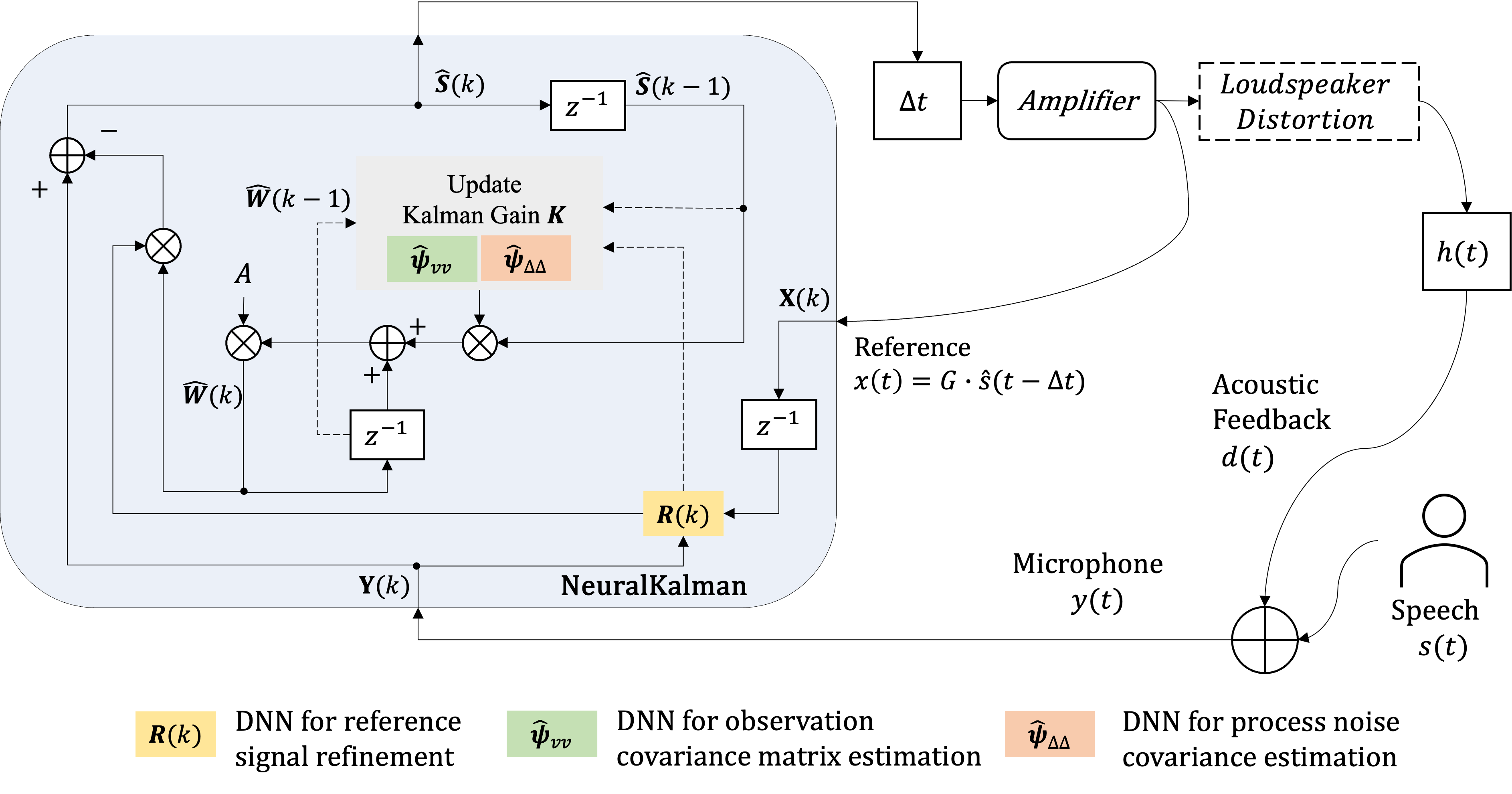}
      \caption{Diagram of an acoustic amplification system and the proposed NeuralKalmanAHS model.}
      \label{fig:neuralkalman}
\end{figure*}

\section{Acoustic Howling Suppression}
\label{sec:ahs}

\subsection{Problem formulation}

% (Refer to previous papers (DeepAHS and HybridAHS) to describe how acoustic echo generated, briefly describe its difference from AEC)

Fig. \ref{fig:neuralkalman} shows a typical single-channel acoustic amplification system with the proposed NeuralKalmanAHS model. The system consists of a microphone and a loudspeaker, both of which are in the same space. The microphone receives a mixture of the near-end speech signal $s(t)$ as well as the playback signal $d(t)$ and sends the mixed signal $y(t)$ to an AHS system for howling suppression. The output from the AHS system is then amplified based on the designated loudspeaker gain and the amplified output $x(t)$ is played out by the loudspeaker.
The playback signal $d(t)$ originated from $x(t)$ can be formulated as,
\begin{equation}
    d(t) = x(t)*h(t),
\end{equation}
where $*$ and $h(t)$ denote convolution and the acoustic path from the loudspeaker to the microphone respectively.
% $NL[\cdot]$ represents the non-linear distortion introduced by the loudspeaker,

Without AHS system, the microphone signal can be formulated as,
\begin{equation}
    y(t) = s(t) + [G \cdot y(t-\Delta t)] * h(t),
    \label{eq:AH}
\end{equation}
where $G$ is the loudspeaker gain and $\Delta t$ denotes the delay between the microphone and the loudspeaker introduced by the system. The recursive relation in Eq.\ref{eq:AH} leads to a re-amplifying of the playback signal which results in acoustic howling - a high-pitched jarring sound.

With AHS system, the formulation becomes,
\begin{equation}
    y(t) = s(t) + [G \cdot \hat{s}(t-\Delta t)] * h(t),
\end{equation}
where $\hat{s}(t)$ is the output from the AHS system. Ideally, the estimated $\hat{s}(t)$ will be as close as $s(t)$. Since acoustic howling is a recursive process, the robustness of the AHS system is decided by how thoroughly the howling sound can be removed in each iteration. 

While acoustic howling and acoustic echo share origins in feedback within communication systems, it is worth noting that they represent distinct issues for two reasons. First, while both stem from playback signals, howling involves recursively accumulated and re-amplified playback signals. Second, in acoustic howling scenarios, the playback signal originates from the same near-end speaker, thereby making AHS more challenging.

%A great idea is to show the different equations for formulating acoustic howling. List the equations for the microphone when "no AHS", "with perfect AHS", "with AHS". And explain the differences among these signals and methods. For models trained with the first two signals, there will be mismatch problem because the realistic signals model received during inference would be the third one.

\subsection{Frequency-domain Kalman filter}
\label{ssec:kalmanfilter}

% (Review the algorithm of Kalman filter based AFC. Describe the details of this method and the limitations it has. Refer to NueralKalman based AEC paper).

Frequency-domain Kalman filter (FDKF) estimates the feedback signal by modeling the acoustic path with an adaptive filter $\mathbf{W}(k)$ ($k$ denotes the frame index). FDKF can be understood as a two-step process, where the iterative feedback from these steps drives the update of filter weights. 

In the prediction phase, the frequency-domain near-end signal $\mathbf{S}(k)$ is estimated by, 
\begin{equation}
    \hat{\mathbf{S}}(k) = \mathbf{Y}(k) - \mathbf{X}(k)\hat{\mathbf{W}}(k),
\end{equation}
where $\mathbf{X}(k)$ is the frequency-domain reference signal which corresponds to the amplified and delayed estimates from the AHS system, $\mathbf{Y}(k)$ corresponds to the frequency-domain microphone signal. $\hat{\mathbf{W}}(k)$ denotes the estimated acoustic path in the frequency domain. 

In the update step, the state equation for updating echo path $\hat{\mathbf{W}}(k)$ is defined as,
\begin{equation}
    \label{eq:state}
    \hat{\mathbf{W}}(k+1) = A[\hat{\mathbf{W}}(k)+\mathbf{K}(k)\hat{\mathbf{S}}(k)],
\end{equation}
where $A$ is the transition factor, and $\mathbf{K}(k)$ denotes the Kalman gain. As shown in Fig. \ref{fig:neuralkalman}, the update of $\mathbf{K}(k)$ is related to the reference signal, acoustic path, and the estimated near-end signal. The dashed line in Fig. \ref{fig:neuralkalman} indicates the relations not expressed directly in the equations. The calculation of $\mathbf{K}(k)$ is defined as,
\begin{equation}
    \mathbf{K}(k) = \mathbf{P}(k)\mathbf{X}^H(k)[\mathbf{X}(k)\mathbf{P}(k)\mathbf{X}^H(k)+\mathbf{\Psi}_{vv}(k)]^{-1},
\end{equation}
\begin{equation}
    \mathbf{P}(k+1) = A^2[\mathbf{I}-\alpha\mathbf{K}(k)\mathbf{X}(k)]\mathbf{P}(k) + \mathbf{\Psi}_{\Delta\Delta}(k),
\end{equation}
 where $\mathbf{\Psi}_{vv}(k)$ and $\mathbf{\Psi}_{\Delta\Delta}(k)$ are observation noise covariance and process noise covariance respectively, $\mathbf{P}(k)$ is the state estimation error covariance. In FDKF, $\mathbf{\Psi}_{vv}(k)$ and $\mathbf{\Psi}_{\Delta\Delta}(k)$ are approximated by the covariance of the estimated near-end signal $\mathbf{\Psi}_{\hat{s}\hat{s}}(k)$ and the acoustic-path $\mathbf{\Psi}_{\hat{W}\hat{W}}(k)$, respectively. More details can be found in \cite{yang2017frequency, enzner2006frequency}.

% \cite{yang2017frequency, enzner2006frequency}

\section{Proposed Method}
\label{sec:method}

\subsection{Overall approach}
% Describe the importance of accurate QR estimation and proper reference signal for Kalman filter. The overall structure of the proposed method.

The overall structure of the proposed NeuralKalmanAHS method is depicted in Fig. \ref{fig:neuralkalman}, where the frequency-domain Kalman filter (described in Section \ref{ssec:kalmanfilter}) is enhanced by integrating NN components to refine the reference signal $\mathbf{X}(k)$ and covariance matrices $\mathbf{\Psi}_{vv}(k)$ and $\mathbf{\Psi}_{\Delta\Delta}(k)$.

\subsubsection{Modeling reference signal $\mathbf{R}$}
As a promising strategy to enhance performance, refining the original reference signal by incorporating a learned reference signal has been established in prior acoustic echo cancellation research \cite{zhang2023hybrid, zhang2023kalmannet} to enhance adaptive algorithm capabilities. To further develop this idea in AHS, we propose to integrate a learned reference signal $\mathbf{R}(k)$ into the Kalman filter framework, with the original reference signal $\mathbf{X}(k-1)$ and microphone recording $\mathbf{Y}(k)$ as inputs:
\begin{equation}
\label{eq:state}
\mathbf{R}(k) = \mathcal{H}_r(\mathbf{Y}(k), \mathbf{X}(k-1)),
\end{equation}
where $\mathcal{H}_r$ represents the network parameters for reference signal estimation.
$\mathcal{H}_r$ is designed as a two-layer long short-term memory (LSTM) network with 300 units per layer followed by a linear layer with Sigmoid activation function, which takes the concatenation of the log power spectrums of the original reference signal and microphone recording as input and estimates a ratio mask which is then applied on the microphone signal to get the refined reference signal $\mathbf{R}(k)$. By integrating this refined reference signal, the operational load on the Kalman filter, particularly in attenuating severe acoustic howling, is reduced, enhancing its efficiency. 
% Traditional algorithms simplify the playback signal as a linear transformation of the reference signal, neglecting nonlinear distortions introduced by amplifiers and speakers. This advancement leverages the capacity of the learned reference signal to encapsulate complex aspects of the acoustic environment, leading to enhanced howling suppression and system robustness.

\subsubsection{Modeling covariance matrices $\mathbf{\Psi}_{vv}$ and $\mathbf{\Psi}_{\Delta\Delta}$}
In the Kalman filter, covariance matrices $\mathbf{\Psi}_{vv}(k)$ and $\mathbf{\Psi}_{\Delta\Delta}(k)$ represent uncertainties associated with measurement and state variables. The accuracy of covariance matrices estimation significantly influences Kalman filter performance, affecting state estimation accuracy, adaptation to dynamic conditions, and convergence rate, which is especially crucial for dependable acoustic howling suppression. Conventional methods for covariance matrix estimation in the Kalman filter often assume linearity and stationary conditions, neglecting variable interdependencies and being sensitive to noise and outliers, limiting adaptability and prediction accuracy. We propose employing NN modules to learn covariance matrices. Original estimations in equations (\ref{eq:Phi_vv}) and (\ref{eq:Phi_dd}) are transformed to:
\begin{equation}
    \label{eq:Phi_vv}
\mathbf{\Psi}_{vv}(k) = \mathcal{H}_{\Psi1}(\mathbf{\hat{S}}(k)),
\end{equation}
\begin{equation}
    \label{eq:Phi_dd}
\mathbf{\Psi}_{\Delta\Delta}(k) = \mathcal{H}_{\Psi2}(\hat{\mathbf{W}}(k)),
\end{equation}
where the estimation of $\mathbf{\Psi}_{vv}(k)$ and $\mathbf{\Psi}_{\Delta\Delta}(k)$ both involve training an LSTM cell with 65 hidden states. The inputs to the RNNs for estimating $\mathbf{\Psi}_{vv}(k)$ and $\mathbf{\Psi}_{\Delta\Delta}(k)$ are the magnitude of estimated near-end speech $\mathbf{\hat{S}}(k)$ and $\hat{\mathbf{W}}(k)$, respectively.
% Using neural networks to estimate these covariance matrices offers distinct advantages, capturing intricate relationships, enhancing accuracy and robustness, and enabling accurate estimation in challenging conditions.

% \subsection{Howling detection during streaming training}

% We observe that training the NeuralKalman model can be difficult and severe acoustic howling is prone to occur with an initially randomized model. The recursive nature of streaming training often results in an energy explosion, leading to a 'not a number' (NAN) issue and halting gradient updates. In the process of training NeuralKalman, we incorporate howling detection as a key measure. Specifically, the output of NeuralKalman is actively monitored during each iteration, with the process halting upon detecting howling. This strategy prevents the recursive training from triggering NAN issues, consequently averting gradient update failures and enhancing the model's convergence.

% \subsection{Estimating covariance matrices: NeuralKalman-QR}
% (Better estimation)

% \subsection{Estimating reference signal: NeuralKalman-f}
% (Modified reference signal for better AHS)

\subsection{Loss function and training strategy}
The loss function in Eq. \ref{eq:loss} relies on the L1 norm to quantify the difference in magnitude spectrum between the enhanced signal $\hat{S}$ and the target signal $S$. By utilizing the L1 loss on the magnitude spectrum, the model benefits from effective regularization of the scale of the output signal. 
% This approach facilitates accurate comparison between the enhanced and target signals, aiding the acoustic howling suppression system in achieving improved performance and preserving the overall quality of the output audio.
\begin{equation}
\label{eq:loss}
    Loss = l1(S, \hat{S})
\end{equation}

We observe that training the NeuralKalmanAHS model can be difficult and severe acoustic howling is prone to occur with an initially randomized model. The recursive nature of streaming training often results in an energy explosion, leading to a `not a number' (NAN) issue and halting gradient updates. In the process of training NeuralKalmanAHS, we incorporate howling detection as a key measure. During each training iteration, we monitor the NeuralKalmanAHS output that utilizes a normalized scale of about -1.0 to +1.0 to interpret the 16-bit WAV file amplitudes. If the amplitude surpasses the upper limit for over 100 consecutive samples—a threshold set from experimental observations—training is halted to prevent howling. This strategy prevents the recursive training from triggering NAN issues, consequently preventing gradient update failures and enhancing the convergence of the model. The proposed model is trained for 60 epochs with a batch size of 128.

\section{Experimental Setup} 
\label{sec:exp}

\begin{table}[]
\centering
\caption{Ablation study on NeuralKalmanAHS components for acoustic howling suppression. Mean and standard deviation are included for all evaluation metrics.}
\begin{tabular}{c|cc}
\specialrule{.15em}{.1em}{.1em}
$G$ = 2       & SDR (dB)                      & PESQ                     \\ \hline
NeuralKalmanAHS & \textbf{2.32 $\pm$ 1.92} & \textbf{2.27 $\pm$ 0.46} \\
without $\mathbf{R}$           & 1.28 $\pm$ 1.42          & 1.72 $\pm$ 0.38          \\
without $\mathbf{\Psi}_{vv} $, $ \mathbf{\Psi}_{\Delta\Delta}$         & 2.17 $\pm$ 1.85          & 2.21 $\pm$ 0.47          \\
Kalman filter       & -11.92 $\pm$ 15.62       & 1.62 $\pm$ 0.80          \\ \specialrule{.15em}{.1em}{.1em}
\end{tabular}
\label{tbl:ablation}
\end{table}

\begin{figure}[]
\centering
\includegraphics[width=5.5cm]{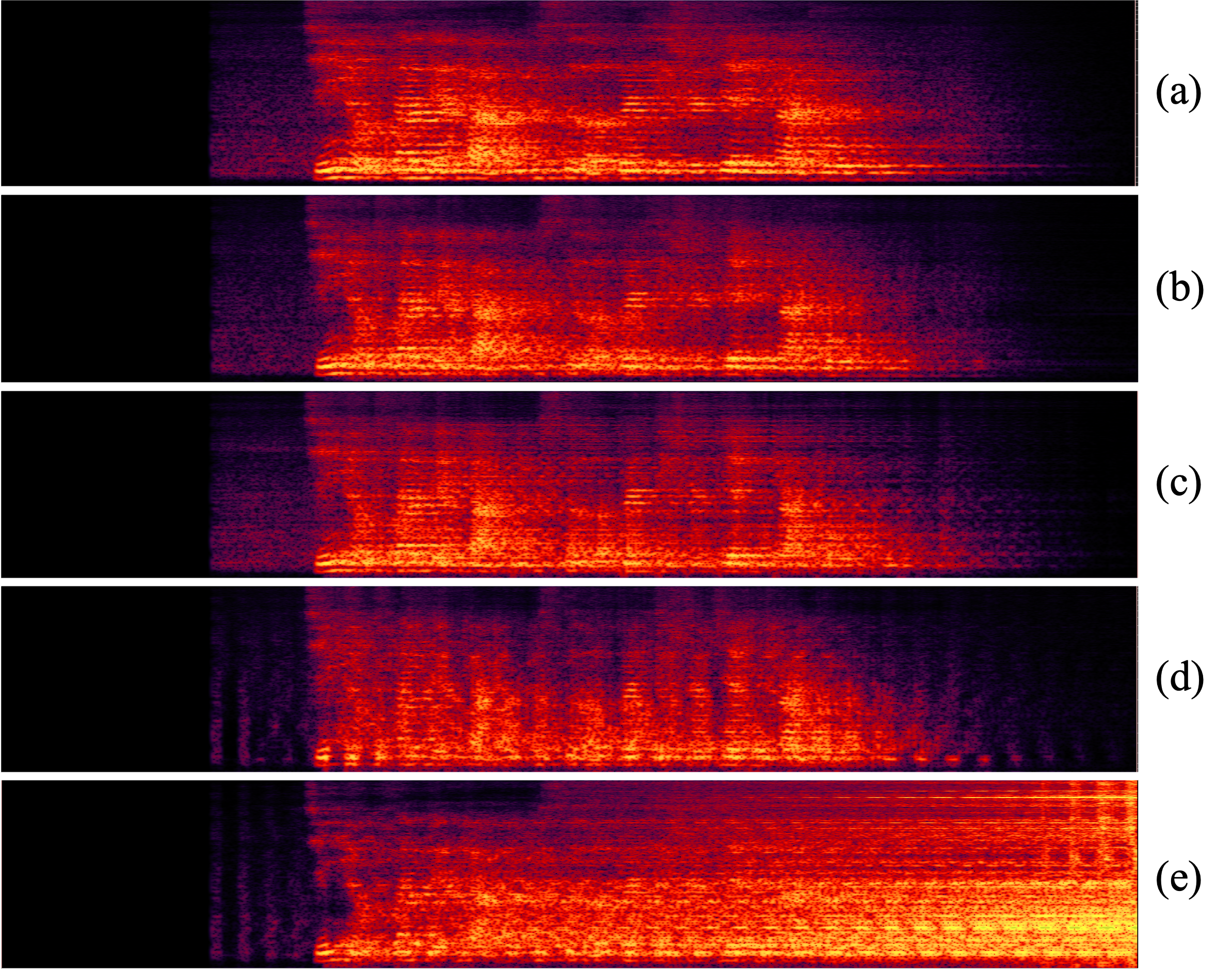}
\caption{Spectrograms of (a) target near-end signal and outputs of (b) NeuralKalmanAHS, (c) NeuralKalmanAHS without $\mathbf{\Psi}_{vv}(k) $, $ \mathbf{\Psi}_{\Delta\Delta}(k)$, (d) NeuralKalmanAHS without $\mathbf{R}(k)$, (e) Kalman filter. }
\label{fig:ablation}
\end{figure}

\begin{table*}[]
\centering
\scriptsize
\caption{Howling suppression performance of different methods. Mean and standard deviation are included for all evaluation metrics.}
\begin{tabular}{c|cccc|cccc}
\specialrule{.15em}{.1em}{.1em}
Models        & \multicolumn{4}{c|}{SDR (dB)} & \multicolumn{4}{c}{PESQ} \\ \hline
G             & 1.5   & 2   & 2.5   & 3  & 1.5   & 2   & 2.5   & 3  \\ \hline
no AHS        &   -30.51 $\pm$ 7.23    &  -31.86 $\pm$ 5.66   &  -33.10 $\pm$ 3.96  &  -33.21 $\pm$ 3.94  & -      &  -   &  -     & -   \\ \hline
Kalman        &   -5.11 $\pm$ 13.20   &  -10.33 $\pm$ 14.84   &   -14.88 $\pm$ 15.14    &  
-18.25 $\pm$ 14.77  &  
1.94 $\pm$ 0.72     &   1.65 $\pm$ 0.73  &    1.44 $\pm$ 0.70   &  1.30 $\pm$ 0.64  \\
DeepMFC \cite{zheng2022deep}        &    -0.09 $\pm$ 6.50   &   -2.78 $\pm$ 9.44  &   -5.59 $\pm$ 11.40    &  
-7.69 $\pm$ 12.26  &   2.11 $\pm$ 0.51    &   1.88 $\pm$ 0.59  &   1.70 $\pm$ 0.62    &  1.56 $\pm$ 0.59  \\
HybridAHS \cite{zhang2023hybrid}     &   2.96 $\pm$ 3.04    &  1.25 $\pm$ 5.79   &   -1.45 $\pm$ 9.60    &  -3.49 $\pm$ 10.90
  &    \textbf{2.57 $\pm$ 0.47}   &  2.33 $\pm$ 0.53   &    
\textbf{2.22 $\pm$ 0.59}   &  1.95 $\pm$ 0.62  \\ 
Neural-KG  &     2.50 $\pm$ 2.78  &   1.63 $\pm$ 3.34  &   -0.46 $\pm$ 7.46
    &  -2.50 $\pm$ 9.94  &    2.35 $\pm$ 0.46   &  2.14 $\pm$ 0.44   &   1.95 $\pm$ 0.48    & 1.80 $\pm$ 0.53   \\ \hline
% NeuralKalman-QR  &       &     &       &    &       &     &       &    \\
% NeuralKalman-f   &   3.35 $\pm$ 2.05    &   2.28 $\pm$ 1.76  &      1.55 $\pm$ 1.60 & 1.04 $\pm$ 1.44   &    2.47 $\pm$ 0.45   &   2.24 $\pm$ 0.42  &    2.07 $\pm$ 0.39   &  1.94 $\pm$ 0.37  \\
NeuralKalmanAHS &   \textbf{3.65 $\pm$ 2.01}    &  \textbf{2.65 
$\pm$ 1.70}   &    \textbf{1.98 $\pm$ 1.49}   &  \textbf{1.45 $\pm$ 1.31}  &    2.55 $\pm$ 0.44   &    \textbf{2.33 $\pm$ 0.41} &   2.17 $\pm$ 0.39    &  \textbf{2.04 $\pm$ 0.37}  \\ \specialrule{.15em}{.1em}{.1em}
\end{tabular}
\label{tbl:comparison}
\end{table*}

\subsection{Data preparation} 

During streaming training and inference, for each sample, a pair of RIRs and speech signals are randomly selected and recursively used to generate the playback and microphone signal. The near-end speech audios are obtained from the AISHELL-2 dataset \cite{du2018aishell}. In addition, 10,000 pairs of room impulse responses (RIRs) are generated using the image method \cite{allen1979image}. The RIRs are characterized by random room properties and reverberation times (RT60) selected within the 0 to 0.6 seconds range. Each RIR pair represents near-end speaker and loudspeaker positions. The system delay ($\Delta$t) of the simulated acoustic amplification system ranges randomly from 0.15 to 0.25 seconds, and the amplification gain is randomly chosen between 1 and 3.
Overall, the training, validation, and testing sets comprised 38,000, 1,000, and 200 utterances, respectively. The testing data contained distinct utterances and RIRs compared to the training and validation data. All input audios are sampled at
16 kHz. STFT is computed with an 8 ms frame length and 50\% frame shift.

\subsection{Evaluation metrics} 

In this study, we evaluate the acoustic howling suppression (AHS) techniques using two metrics: signal-to-distortion ratio (SDR) \cite{vincent2006performance} and perceptual evaluation of speech quality (PESQ) \cite{rix2001perceptual}. While we rely on PESQ to evaluate speech quality preservation, we emphasize SDR results to show the effectiveness of AHS methods in howling suppression, considering the insensitivity of PESQ to scale.

% We emphasize SDR to measure how effectively AHS reduces howling artifacts, while PESQ can assess speech quality preservation given its insensitivity to scale.

\section{Experimental Results}
\label{sec:results}

\subsection{Ablation study}

% Kalman only, without f
% f only, without Kalman
% With kalman, but use f's output
% With kalman, but use f's input
% With kalman, use kalman's output
% Check whether NN could learn the delay
% Make NeuralKalman-QR the same size as NeuarlKalman-f
% 

We perform an ablation study on NeuralKalmanAHS, evaluating the role of modeling reference signal and covariance matrices. Using 50 utterances randomly selected from the test set and evaluating with a fixed loudspeaker gain of 2, we train and compare various model versions: complete NeuralKalmanAHS, NeuralKalmanAHS without modeling covariance matrices, NeuralKalmanAHS without modeling reference signal, and original Kalman filter. 
Table \ref{tbl:ablation} demonstrates the significance of modeling both $\mathbf{R}(k)$ and $\mathbf{\Psi}_{vv}(k)$, $ \mathbf{\Psi}_{\Delta\Delta}(k)$ components. We can observe that although not modeling $\mathbf{R}(k)$ results in a model only estimating covariance matrices for having just 0.08 M parameters and inadequate performance in terms of speech quality, with streaming training, the model still shows robustness in preventing severe howling, achieving significantly higher mean and lower standard deviation of SDR compared to the Kalman filter. 
In addition, we observe that only estimating $\mathbf{R}(k)$ shows strong howling suppression performance, SDR and PESQ are slightly worse compared to the complete NeuralKalmanAHS.  
Spectrograms in Fig. \ref{fig:ablation} which visualize differences among evaluated models in howling suppression also validate these observations. 
% \textsuperscript{\ref{note:demo}}

% \footnote{\label{note:demo}Demos are provided in \url{https://yixuanz.github.io/NeuralKalmanAHS}.}

\subsection{Comparison with other methods}

% Final results. Compare with other baseline methods: Offline methods, G-voice/ICASSP paper.

\begin{figure}[]
\centering
\includegraphics[width=8.6cm]{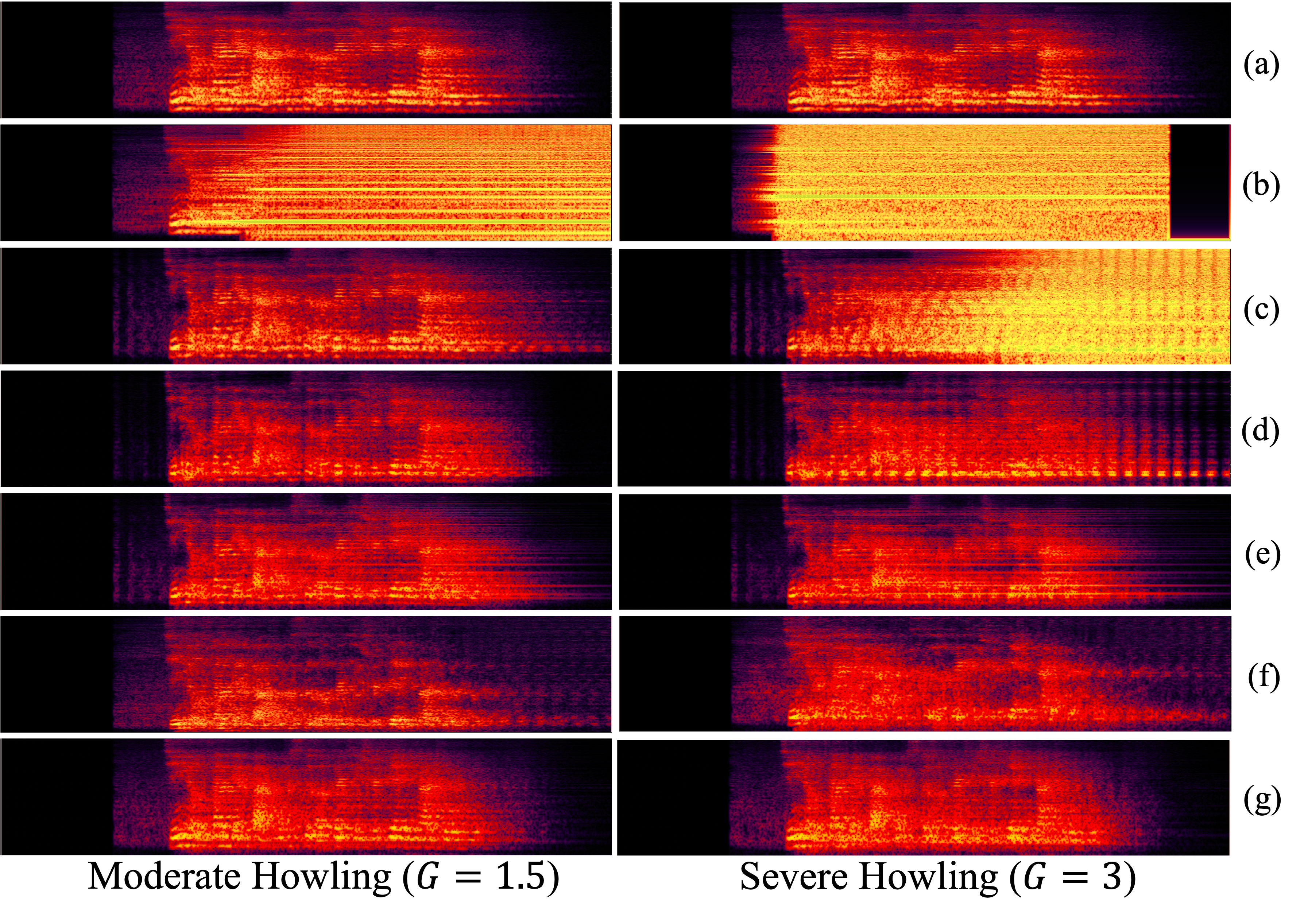}
\caption{Spectrograms of (a) target signal, (b) no AHS, (c)
Kalman filter, (d) DeepMFC, (e) HybridAHS, (f) Neural-KG, and (g) Proposed NeuralKalmanAHS.}
\label{fig:comparison}
\end{figure}

Table \ref{tbl:comparison} compares NeuralKalmanAHS to several benchmarks:  the frequency-domain Kalman filter, two offline-trained methods including DeepMFC \cite{zheng2022deep} and HybridAHS \cite{zhang2023hybrid}, and an online-trained model Neural-KG. The Neural-KG, inspired by \cite{soleimani2023neural}, modeling Kalman gain in the Kalman filter, is built upon \cite{yang2023low} and uses streaming training. For a fair comparison, all NN-based methods, unless explicitly mentioned, utilize a two-layer LSTM network with a model size and experimental settings aligned with NeuralKalmanAHS. We evaluate the models at loudspeaker gains ($G$) of \{1.5, 2, 2.5, 3\}, where lower gains imply less challenging scenarios.
Results show that DNN-based methods consistently outperform the frequency-domain Kalman filter in all cases. While DeepMFC and HybridAHS perform well at low $G$,  their efficacy diminishes as $G$ rises. On the other hand, NeuralKalmanAHS remains robust in challenging situations with high loudspeaker gain values.  Compared to the best-performed HybridAHS, NeuralKalmanAHS demonstrates robust howling suppression performance with reduced distortion, especially in challenging cases. When $G=3$, it enhances SDR by 4.94 dB and PESQ by 0.09. Spectrograms of the estimates from all benchmark methods are illustrated in Fig. \ref{fig:comparison}. We can observe that the proposed NeuralKalmanAHS consistently achieves the best results across both moderate and severe howling scenarios.

\section{Conclusion}
\label{sec:conclusion}
In this study, we have introduced NeuralKalmanAHS, an NN-augmented Kalman filter for acoustic howling suppression. Our approach employs NN to help refine reference signal and estimate covariance matrices in the frequency-domain Kalman filter. Through an ablation study, we have demonstrated the significance of modeling the covariance matrices and reference signal, and the efficacy of streaming training, even when focusing solely on modeling covariance with a compact model. The proposed NeuralKalmanAHS outperforms strong DNN-based benchmarks and exhibits less distortion. Future work will explore lightweight model designs and extend the approach to multi-channel acoustic howling suppression.

% Our approach incorporates a howling detection strategy, allowing effective training in streaming mode.

\vfill\pagebreak

\bibliographystyle{IEEEbib}
\bibliography{strings,refs}

\end{document}